\documentclass[a4paper,twocolumn,10pt]{article}
\pdfoutput=1
\usepackage{graphicx}
\usepackage{ucs}
\usepackage[utf8x]{inputenc}
\usepackage[pdftex,colorlinks=false,pdfborder={0 0 0.2},urlbordercolor={0.2 0.2 1}]{hyperref}
\usepackage{color}
\usepackage{colortbl}
\usepackage{longtable}
\usepackage{calc}
  {\begin{list}{}%
       {\setlength\rightmargin{\rightmargin-3.5cm}%
        \setlength\leftmargin{\rightmargin}}%
       \item[]
  }%
  {\end{list}}%

\usepackage{pslatex}

\makeatletter
\def\maxwidth{%
  \ifdim\Gin@nat@width>\linewidth
    \linewidth
  \else
    \Gin@nat@width
  \fi
}
\makeatother

\def\tablecaption{}
\def\imagecaption{}
\def\pretable{}
\def\posttable{}
\def\imagewidth{\maxwidth}

\def\wikititle{Evaluating Trust in Grid Certificates}

\begin{document}
%

\title{\wikititle \thanks{ Accepted for SAC'10 March 22--26, 2010, Sierre, Switzerland.} }

\author{David~O'Callaghan, Louise~Doran and Brian~Coghlan\\ Trinity College Dublin\\ Dublin 2, Ireland\\ david.ocallaghan@cs.tcd.ie, doranlr@tcd.ie, coghlan@cs.tcd.ie}

\maketitle

\begin{abstract} Digital certificates are used to secure international computation and data storage grids used for e-Science projects, like the Worldwide Large Hadron Collider Computing Grid. The International Grid Trust Federation has defined the Grid Certificate Profile: a set of guidelines for digital certificates used for grid authentication. We have designed and implemented a program and related test suites for checking X.509 certificates against the certificate profiles and policies relevant for use on the Grid. The result is a practical tool that assists implementers and users of public key infrastructures to reach appropriate trust decisions. \end{abstract} 



\section{Introduction}
\label{sec:Introduction} 

The international e-Science grid community is heavily dependent on X.509 \cite{X.509} digital certificates issued by certification authorities (CAs). 

The International Grid Trust Federation, working through the Open Grid Forum, has defined the Grid Certificate Profile as a set of guidelines for digital certificates used for grid authentication \footnote{See \href{http://igtf.net/}{http://igtf.net/} and \href{http://ogf.org/}{http://ogf.org/}. 

}. 

A {\em relying party}, i.e. anyone who must trust the public key infrastructure in order to make use of the system, may wish to check that a CA is compliant with the Grid Certificate Profile, or some other set of requirements. This cohort would also include grid system administrators or security auditors acting on behalf of grid users. 

The process to check that a digital certificate is compliant is largely manual: the relying party must work their way through the Grid Certificate Profile section-by-setion checking that the certificate under examination meets the requirements of each provision. This would typically require the certificate to be output in a human-readable form and also in a representation of the underlying ASN.1 structure.

As the certificates are computer-readable and the Grid Certificate Profile is relatively well-defined, there is clearly scope for automation of this compliance check. This should make it easier and more accurate to check that a CA is complying with requirements, and this should allow relying parties to check compliance as necessary, rather than relying on informal assertions made by others. It should also allow CAs implementing PKI for the grid to check that they are compliant before applying for accreditation as grid CAs, thereby reducing effort. 

This is particularly important as the number of grid CAs continues to grow as new communities join. In parallel, use of the grid -- and with it the importance of the trust infrastructure -- is increasing. For these reasons, we believe that a tool to automate certificate checking is important for scalability. 

\section{Background}
The {\em European Policy Management Authority for Grid Authentication in e-Science} (EU Grid PMA) \footnote{See \href{http://www.eugridpma.org/}{http://www.eugridpma.org/} 

} was established in 2004 to develop practices, requirements and policies for CA coodination in Europe and internationally, emerging from work done in earlier European grid projects to create a large international X.509 public key infrastructure for grid authentication \cite{astalos.05.egc}.  {\em The Americas Grid Policy Management Authority} and the {\em Asia-Pacific Grid Policy Management Authority} were established on a similar basis to coordinate grid authentication policies in those regions\footnote{See \href{http://tagpma.org/}{http://tagpma.org/} and \href{http://www.apgridpma.org/}{http://www.apgridpma.org/} respectively 

}, and the {\em International Grid Trust Federation} (IGTF) was established in October 2005 to coordinate policies and practices between the regional grid authentication policy management authorities \cite{igtf.05.nr}.  The Open Grid Forum has a CA Operations working group that works closely with the IGTF to establish policies and procedures.  

The conventional approach to building a large-scale X.509 PKI is to set up a hierarchy of certification authorities with a single root CA and a number of sub-CAs and sub-sub-CAs with various assurance levels, purposes, catchment areas, etc. For technical and organizational reasons, the model adopted for grid authentication can be characterised as a {\em policy-based bridge} where each CA meeting a policy is accredited by a PMA and a relying party who trusts the PMA can trust the accredited CAs. 

The grid PMAs have developed {\em Authentication Profiles} for different classes of certification authority which cover many aspects of grid CA operations. 

One important aspect of CA policy is the {\em certificate profile} used when issuing a certificate. This defines technical requirements that are fundamental to trust. The IETF PKI working group (PKIX) is developing a standard profile for X.509 certificates for use on the Internet in RFC 5280 \cite{rfc5280}. Building on this, the IGTF and the OGF have developed a {\em Grid Certificate Profile} that deals with aspects of the profile that are of particular importance in the context of grid authentication and trust \cite{GFD.125}.  

The idea of automated checking of grid certificates and profiles has been a topic of discussion in EU Grid PMA and its predecessor, the European DataGrid CA coordination group, since 2001. Coghlan presented a system for checking CA certificate policies and certification practice statements against requirements in 2002 \cite{Coghlan:02:GGF}. O'Callaghan and Coghlan developed a system in 2006 for evaluating trust in Grid CAs which uses information both from CA policies and certificates \cite{OCallaghan:06:Grid}.  

The NIST Computer Security Division have documented a suite of tests for certificate compliance with RFC 3280 (now obsoleted by RFC 5280, mentioned above)\footnote{See \href{http://csrc.nist.gov/groups/ST/crypto_apps_infra/pki/pkitesting.html}{http://csrc.nist.gov/groups/ST/crypto\_apps\_infra/pki/pkitesting.html} 

}. 

It seems possible that other PKI implementers have taken a similar approach to check that a certificate meets specified requirements or is compatible with a given profile. However, these tools do not appear to be generally available (on the web) or to be documented in the literature. 

\section{Design}
The main aim was to design a practical tool for use in the e-Science grid context. The target users are CA operators, CA auditors, and system administrators who wish to check for themselves that a CA is operating according to their policy.  

\subsection{Languages and Libraries}
To implement the tool a number of programming languages and their associated libraries were considered. The initial candidates were Kawa Scheme, Java and Perl.  

Kawa is an implementation of Scheme on the Java Virtual Machine \cite{bothner:freenix98}. Kawa was considered as earlier trust evaluation software was implemented in this environment \cite{OCallaghan:06:Grid}. Kawa can use Java libraries and the earlier software used BouncyCastle for functions related to cryptography and certificates \cite{BouncyCastle}. This candidate suffers from the relative obscurity and unfamiliarity of the language among the target users. 

Java was considered as it is widely available and, again, has access to BouncyCastle as a cryptography and certificate library. However, despite its widespread use it was not considered suitable as it is not a scripting language. 

Perl and Python were also considered as scripting languages widely used by system administrators. In IGTF discussions, Perl was favoured as it met the perceived technical requirements of the target user community. 

To work alongside Perl, the main candidate cryptography and certificate library was OpenSSL. The library is used by major Grid middleware, such as Globus \cite{Foster:05:ICNPC}. The OpenSSL command-line tool is often used for PKI administration tasks; and many accredited grid CA services use OpenSSL at their core. For these reasons OpenSSL was chosen. 

There are a number of available Perl bindings for access to the OpenSSL library and/or command: the candidate modules considered were OpenCA::\-X509 and Crypt::\-OpenSSL::\-X509\footnote{Other modules which were not considered at the time but which may merit investigation are Crypt::X509 and Convert::X509. Both of these modules are available through CPAN (\href{http://cpan.org}{http://cpan.org}). 

}. 

The OpenCA::\-X509 module was developed in the OpenCA PKI project\footnote{See \href{http://openca.org/}{http://openca.org/} 

} and was considered because a number of grid CAs successfully use this software. It provides access to OpenSSL through the OpenSSL command-line tool (that is, API calls result in spawned OpenSSL command processes). This has the disadvantage of making the software dependent on the human-readable output of the command-line. The module appears to be dependent on other modules and libraries distributed with the OpenCA package, which would tend to make it impractical for use as a stand-alone tool for systems administrators. 

The Crypt::\-OpenSSL::\-X509 module provides access to some of the certificate-handling ability found in OpenSSL. It interfaces directly with the OpenSSL library API. In addition, the module is available in software repositories for popular Linux OS distributions (e.g. Fedora and Ubuntu) making it readily available for the intended users. The disadvantage of the Crypt::\-OpenSSL::\-X509 module is that it does not provide access to X.509 certificate extensions and some other aspects of the OpenSSL API. This module was chosen as the basis for certificate handling due to its availability and the possibility that improvements to functionality (described below) might be distributed by Linux vendors in future. 

\subsection{Test Framework}
The requirements on the design of the tool were to provide a way to specify assertions and comparisons of certificates against profiles; and to provide a framework to evaluate a certificate against a suite of these tests and get a pass/fail result, a trust score, or advice on possible problems. 

The first step was to design the format for the test suites. Ideally, a test format should be portable so that different implementations can share and re-use tests. A portable test format in XML, S-expressions, or JSON\footnote{See \href{http://json.org/}{http://json.org/} 

} was considered. The disadvantage was that in each case a language would need to be defined for assertions and comparisons, and for access to certificate fields and attributes.  

As Perl had been chosen as the de facto implementation language, an alternative approach is to structure the tests as fragments of Perl code arranged in an associative array where each key was the name of a test and each value was a test function. This design is similar to that of O'Callaghan \cite{OCallaghan:06:Grid}. An initial prototype was created according to this design but displayed the disadvantage that it did not directly provide a framework for collating test results. 

A third alternative is to use the standard Perl test framework, which produces output in the Test Anything Protocol (TAP) format\footnote{See \href{http://testanything.org/}{http://testanything.org/} 

}. Many Perl modules include a suite of functional tests written for the Test::\-Harness module using the Test::\-More module. Test::\-More provides a set of commands for assertions and comparisons between expected and computed values: {\em is}, {\em isnt}, {\em like}, {\em unlike}, etc.  Test::\-Harness allows suites of tests to be run and the results collected and presented\footnote{Test::Harness was used rather than the newer TAP::Harness because a version of the former is typically distributed with Perl. 

}. The test format provided by Test::\-More is very much tied to Perl, but is eminently practical given the requirements. 

\subsection{Test Suites}
The test suites for the Grid Certificate Profile are designed to follow the structure of the document as far as possible, with a test for each testable provision. The Profile contains provisions related to different classes of certificates: those belonging to CAs and those belonging to {\em end entities}, i.e. hosts, persons and robots\footnote{where a {\em robot} is an automated client entity that is not a natural person. 

}. It is logical to provide test suites for each class. In addition, the patterns used for test suites should be applicable to other tests, such as security vulnerability checks or checks on certificate revocation lists (CRLs). 

\section{Implementation}
The implementation can be divided into three parts: the {\em checkcerts.pl} script, which runs the tests and collects the results; the test suites, specifically the test suites for the Grid Certificate Profile; and the enhancements made to the Crypt::\-OpenSSL::\-X509 library to allow implementation of the necessary tests. 

\subsection{Test Script}
The central script is a simple wrapper around the Test::\-Harness module. It takes as parameters a list of test files and a list of certificate files (which should be in PEM format). The {\em aggregate} option causes all tests on all certificates to be executed in a single test run to give an overall {\em PASS} or {\em FAIL} result, which may be useful to verify compliance of a set of certificates. 

As the Test::\-Harness module does not provide a mechanism for passing parameters (such as the paths of certificate files) to an individual test or test suite it was necessary to work around this limitation. 

The approach taken was to output the list of certificates to a known file. Each test suite intended for use with {\em checkcerts.pl} must {\tt use} the supplied {\tt CheckCertsTest} module, which will open the known file and store the list of certificates in an array variable available to the test suite. 

\subsection{Test Suites}
A typical Perl test suite contains a series of test commands. To allow multiple certificates to be tested, the certificate test suites are structured as a loop over an array of certificates. The first test in the test suite loads the certificate to be examined from a file; this can be considered a side-effect. 

As the motivating example, the initial effort was directed towards implementing the provisions of the Grid Certificate Profile as a number of test suties, one for each class of certificate: CA certificates, host or server certificates, personal certificates and robot certificates. 

Each test consists of two parts: a comparison and a message. The comparison will typically contain a calculated or retrieved value, an operator and an expected value. For example:

\begin{verbatim}
cmp_ok($x509->version, '==', 2,
  'Version value MUST be "2" per X.509v3 
   (2.1)');
\end{verbatim}
In this example, {\tt cmp\_ok} is the test command: it tests the retrieved value on the left {\tt \$x509-$>$version} (the certificate format version) against the required value {\tt 2} on the right using the equality operator {\tt ==}. The final parameter to the test command is the output message; in this test suite it is the provision stated in Section 2.1 of the Grid Certificate Profile. If the certificate were to fail this test the message would be printed to indicate where the problem lies. 

The test framework also supports regular expression comparisons, for example:

\begin{verbatim}
unlike($x509->sig_alg_name, '/md5/i',
  'Message digest MUST NOT be MD5 in new
   CA certs (2.2)');
\end{verbatim}
That is, the signature algorithm name must not match {\tt md5}. 

In addition to tests on the certificate object ({\tt \$x509} in the examples) the test suites examine the subject name and certificate extensions in more detail. These areas in particular required enhancements to the Crypt::\-OpenSSL::\-X509 module.  

The unmodified module provides the subject name as a string; whereas the profile requires examination of the ASN.1 type, encoding and value of elements within the name. Functions were added to the {\tt X509} object to return an ordered list of name elements; to query the existence of a given type of element in a name; to search for an element by type (by type name or by object identifier); and to query the encoding of name elements (e.g. {\em ia5String}, {\em printable\-String}, etc.). These additions allowed Sections 2.3  and 3.2 to be implemented\footnote{{\em Issuer and Subject names of Certification Authority certificates} and {\em Subject distinguished names of End-entity certificates} respectively 

}. 

The unmodified Crypt::\-OpenSSL::\-X509 module also does not provide general support for access to certificate extensions, but the underlying OpenSSL API provides functions to retrieve extensions by index and also to convert the object identifier (OID) to a readable name. This allowed construction of an associative array relating extension names to values. An example is:

\begin{verbatim}
my $exts = $x509->extensions_by_name();
...
ok($$exts{'basicConstraints'}->critical(),
  'basicConstraints SHOULD be marked critical
  (2.4.1)');
\end{verbatim}
To implement the provisions of Sections 2.4 and 3.3 of the Grid Certificate Profile (dealing with extensions in CA certificates and end-entity certificates respectively) it was necessary to extend Crypt::\-OpenSSL::\-X509 to give it some semantic information for relevant extensions, such as {\em basicConstraints}, {\em keyUsage}, and {\em extendedKey\-Usage}. 

It should be noted that the provisions of the Grid Certificate Profile make use of RFC 2119  terminology (\textsc{must}, \textsc{must not}, \textsc{should}, \textsc{should not}, etc.) for requirements \cite{rfc2119}. The tests defined with the Perl Test modules either {\em PASS} or {\em FAIL}: there are no intermediate degrees. If a certificate does not implement a \textsc{should} provision it will {\em FAIL} that test. This is considered acceptable within IGTF as an unimplemented \textsc{should} provision must be clearly justified to the accrediting PMA. For other uses \textsc{should} provisions could be placed in a separate test suite. 

In addition to the Grid Certificate Profile tests, test suites have been developed to implement security vulnerability checks highlighted by the IGTF Risk Assessment Team (or RAT)\footnote{See  \href{http://tagpma.es.net/wiki/bin/view/IGTF-RAT/}{http://tagpma.es.net/wiki/bin/view/IGTF-RAT/} 

}. In particular, these check certificates for vulnerable RSA parameters, known-weak RSA keys due to a recent Debian vulnerability, and the vulnerable hash algorithm MD5. Finally, a small IGTF RAT test suite against CRLs has been created to check for use of the MD5 hash algorithm. This required further development of the Crypt::\-OpenSSL::\-X509 module to support access to CRL objects. 

\section{Results}
\subsection{Grid Certificate Profile}
The Grid Certificate Profile contains 88 distinct provisions. Of these, 44 pertain to CA certificates and 65 to end-entity certificates; 21 are common to both. Of these 4 pertain only to host certificates while 61 are common to all end-entity certificates, for hosts, persons and robots. 

Of these 88 provisions, 80 were found to be implementable as automatic tests. The remaining 8 are not yet implemented in our system for a number of reasons. 4 require comparing multiple certificates (e.g. to check uniqueness of serial numbers or subject names). One requires an online check: that a {\em cRLDistributionPoint} URI refers to a DER-encoded CRL. A further 3 require a manual check or a value judgement. For example, whether an OCSP service is of production quality would require investigation (from Section 3.3.13).  

In total, 61 tests were implemented. There remain 19 implementable provisions to be completed. 

We have implemented 69 percent coverage of the Grid Certificate Profile and 75 percent of its implementable provisions. It may be the case that some provisions should be given more weight than others. 

\subsection{Compliance of IGTF CAs}
At the time of writing the latest version of the Grid Certificate Profile test suite for CA certificates was run against the CA certificates in the IGTF distribution. This gives an overall picture of profile compliance. It also highlights some short-comings in the test suite. 

Only 22 out of 91 CA certificates fully passed the test suite. Two certicates belonging to one CA (now superceded) failed on one or more \textsc{must} provisions (specifically, the use of the MD5 hash algorithm)  and another 69 failed on one or more \textsc{should} provisions. The most common causes of failure are given in Table 1. 

\def\tablecaption{\label{causes}Common causes of failures.} 

\newlength{\tablewidth}

\ifx\floatsspancolumns\undefined
\begin{table}[htbp]\else\begin{table*}[htbp]\fi\begin{center}\pretable{}\arrayrulecolor{black}\begin{tabular}{ll}
\cline{1-2}{  Cause     }&{ Certificates failing   }\\ 
\cline{1-2}{ Serial number   }&{ 39                     }\\ 
{ {\em nsCertType}    }&{ 32   }\\ 
{ Subject Name       }&{ 22   }\\ 
\cline{1-2}\end{tabular}\caption{\tablecaption}\posttable{}\end{center}\ifx\floatsspancolumns\undefined
\end{table}\else\end{table*}\fi
\def\tablecaption{}\posttable{}

By way of explanation, the serial number should not be equal to zero\footnote{This requirement is additional to the Grid Certificate Profile. 

}; the use of the {\em nsCertType} extension is deprecated and, if used, must be consistent with {\em keyUsage}; and the subject name should have the correct encoding and components. 

\subsection{Compliance over time}
The correlation between compliance and the date when a CA was accredited by a grid PMA was investigated. Date of CA certificate issue, as indicated by the {\em notBefore} attribute, is used as a proxy variable for date of accreditation. In general, the certificate will be issued shortly before accreditation. However, a CA may update its certificate without necessarily updating its certificate profile. A plot of the number of Grid Ceritifcate Profile failures against CA certificate {\em notBefore} date is shown in Figure \ref{plot}. 

\def\floatsspancolumns{} \def\imagecaption{\label{plot}Plot of number of Grid Certificate Profile failures against CA certificate notBefore date.} \ifx\unfloat\undefined\ifx\floatsspancolumns\undefined
\begin{figure}[htbp]\else\begin{figure*}[htbp]\fi\begin{center}\fi
\includegraphics{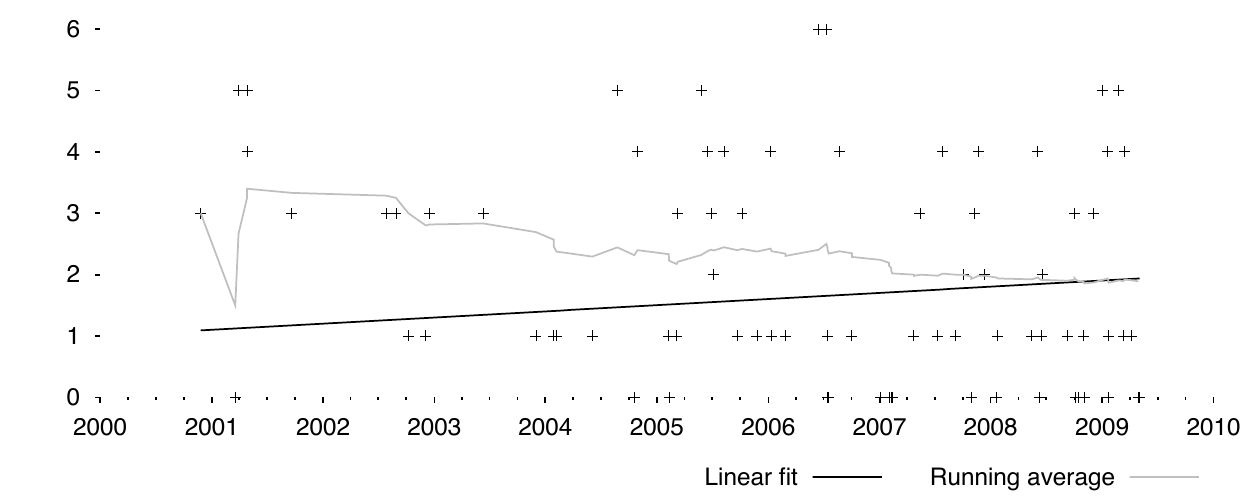}\ifx\unfloat\undefined\caption{\imagecaption}\end{center}\ifx\floatsspancolumns\undefined
\end{figure}\else\end{figure*}\fi\fi
\def\imagecaption{}
\def\imagewidth{\maxwidth}
\let\unfloat\undefined

Overall, a linear fit to the data points indicates a very slight increase in failures (a decrease in compliance) over time. Note that the number of failures ranges from zero to six out of a possible 61. A plot of the average number of failures per CA appears to show a decrease over time. In addition, it appears that the number of compliant CAs is increasing over time but this may simply be due to the increase in the total number of CAs. 

\section{Discussion}
It is encouraging to see that the approach to trust evaluation of certificates presented here has allowed us to implement automated tests for a significant proportion of the provisions of the Grid Certificate Profile; potentially over 90 percent will be implementable. For IGTF relying parties it is reassuring to have the technical means to verify compliance to such a large degree. 

It would be desirable to improve the test suites to clearly indicate if a known provision has not been implemented so that a relying party may make the necessary checks. The implementation of the test suite also highlighted the fact that individual provisions in the Grid Certificate Profile are not unambiguously labelled.  This may be useful input for a subsequent document revision. 

The results for compliance amongst IGTF accredited CAs are disappointing but not entirely unexpected. It should be noted again that the failures were largely against \textsc{should} provisions, i.e. strong recommendations. A CA may be able to justify its lack of compliance. Some of the failures can be explained by virtue of the age of the CA certificate. For example, to quote from Section 2.4.4 on the use of {\em nsCertType} and related attributes: The {\em ns*} attributes are deprecated and \textsc{should not} be included in any new CA certificates. 

The correlation between date of issue of the CA certificate and its compliance is interesting. The results indicate that full compliance cannot be assumed for newly issued certificates. One possible cause is older CAs issuing a new certificate using a non-compliant profile. A correlation between CA software and compliance with the test suite might also be revealing. 

An issue that arose during development of the test suites was the distinction between test assumptions and test provisions. As mentioned, some provisions in the Grid Certificate Profile apply specifically to, say, CA certificates. The test suite for CA certificates is written with the assumption that it will be run against CA certificates. If the CA certificate test suite is passed a certificate with, for example, {\em basicConstraints} attributes that are not appropriate for a CA certificate this will be indicated as a specific test failure. 

The test suite for the Grid Certificate Profile does not address all aspects of grid CA compliance with the IGTF accreditation requirements. For example, the IGTF Authentication Profile for Classic X.509 CAs \cite{EUGridPMA.08.Classic} has requirements for ceritificate validity period and key lengths. In addition, it imposes requirements on CA revocation services. 

The technical certificate profile checks provided by the tool described here are a useful complement to automated checks on Certification Practice Statement documents \cite{rfc3647}. One advantage that this software has over, for example, the system (based on Kawa Scheme and Java) presented in \cite{OCallaghan:06:Grid} is that it is more practical and accessible for system administrators. In addition, it has better coverage of certificate profile tests. It would also appear to be readily usable with higher-level tools that can make use of the TAP output from the Perl Test framework. 

\vspace{1cm} 

\section{Conclusions}
A number of items of future work emerge from the results of this project. There is a clear need to get as close to complete coverage of the Grid Certificate Profile as possible. In addition, complementary test suites for other grid and PKI requirements or standards are desirable. 

With reference to the software, it would be useful to have an online system to allow certificates to be uploaded for testing. It might also be possible to use the software as an input to an online certificate status or certificate validation service. In addition, since the distribution of grid CA certs is easily available a web application to present the current test results for all CAs would be attractive. This would be an update to the CA Trust Matrices, presented in \cite{Coghlan:02:GGF}. 

Nevertheless, in its current state the tool and test suites as described comprise a useful practical tool for grid PKI implementers and users, and indeed IGTF reviewers have begun to make use of it when assessing CAs for accreditation. 

\section{Acknowledgements}
The authors would like to thank IGTF members for their feedback. The work was funded in part by the European Commission through EGEE-III (INFSO-RI-222667) and in part by the Irish Higher Education Authority's Programme for Research in Third-Level Institutions through the e-INIS project. Much of the development work was carried out during the TCD School of Computer Science and Statistics Summer Internship Programme 2008-09. 

\raggedright\bibliographystyle{abbrv}\bibliography{PublishedPapersBibtex,CheckCertsPaperBibtex} 

\end{document}